\documentclass[useAMS, twocolumn, usenatbib]{mn2e}
\usepackage[breaklinks,colorlinks,urlcolor=blue,citecolor=blue,linkcolor=blue]{hyperref}
\usepackage{mathrsfs}
\usepackage{graphicx}
\usepackage{amsmath, amssymb}
\usepackage{graphicx}
\usepackage{epsfig}
\usepackage{enumerate}

\usepackage{color}

\newcommand       \be          {\begin{eqnarray}}
\newcommand       \ee          {\end{eqnarray}}

\newcommand\degd{\ifmmode^{\circ}\!\!\!.\,\else$^{\circ}\!\!\!.\,$\fi}

\newcommand{\lsim}{\stackrel{\scriptstyle <}{\scriptstyle \sim}}

\begin{document}

\title[Stellar Mass Black Hole Mergers in AGN Disks]{Assisted Inspirals of Stellar Mass Black Holes Embedded in AGN Disks: Solving the ``Final AU Problem''}
\author[Stone, Metzger, \& Haiman]{Nicholas C.~Stone$^{1}$\footnotemark[1], Brian D.~Metzger$^{1}$, Zolt\'an Haiman$^{1}$
\\$^{1}$Columbia Astrophysics Laboratory, Columbia University, New York, NY, 10027
\\ \footnotemark[1] Einstein Fellow; email nstone@phys.columbia.edu}
\maketitle
%\altaffiltext{1}{Columbia Astrophysics Laboratory, Pupin Hall, New
% York, NY, 10027, USA}
 
 %\altaffiltext{2}{Einstein Fellow; nstone@phys.columbia.edu}

\begin{abstract}
We explore the evolution of stellar mass black hole binaries (BHBs) which are formed in the self-gravitating disks of active galactic nuclei (AGN).  Hardening due to three-body scattering and gaseous drag are effective mechanisms that reduce the semi-major axis of a BHB to radii where gravitational waves take over, on timescales shorter than the typical lifetime of the AGN disk.  Taking observationally-motivated assumptions for the rate of star formation in AGN disks, we find a rate of disk-induced BHB mergers ($\mathcal{R} \sim 3~{\rm yr}^{-1}~{\rm Gpc}^{-3}$, but with large uncertainties) that is comparable with existing estimates of the field rate of BHB mergers, and the approximate BHB merger rate implied by the recent Advanced LIGO detection of GW150914.  BHBs formed thorough this channel will frequently be associated with luminous AGN, which are relatively rare within the sky error regions of future gravitational wave detector arrays.  This channel could also possess a (potentially transient) electromagnetic counterpart due to super-Eddington accretion onto the stellar mass black hole following the merger. 
 \end{abstract}

%\keywords{gamma-ray bursts: general --- radio continuum: general ---
%  supernovae: general --- surveys}

\section{Introduction}
\label{sec:intro}

Binaries consisting of stellar-mass black holes (BHBs) may coalesce due to gravitational wave (GW) emission and are therefore promising sources for detection by the interferometers Advanced LIGO and Virgo (\citealt{Abadie+11}).   Advanced LIGO recently announced the discovery of the gravitational wave event GW150914, shown to be a BHB with initial component masses $M_1 \approx 36M_{\odot}$ and $M_2 \approx 29M_{\odot}$ (\citealt{LIGO+16}).  The Kerr parameter of the primary black hole was constrained to be $a \lesssim 0.7$, while that of the secondary is only weakly constrained.  This exciting discovery has initiated the era of GW astronomy. 

The discovery of a BHB merger so early in the Advanced LIGO experiment implies a relatively high volumetric rate of $\sim 2-400$ Gpc$^{-3}$ yr$^{-1}$ \citep{LIGO+16b}.  This is broadly consistent with the range of rate estimates from population synthesis modeling of field binaries (e.g. \citealt{Voss&Tauris03, Belczynski+10,Dominik+12,Dominik+13}).  For instance, \citet{Dominik+15} predict a rate of $\sim 1-100$ Gpc$^{-3}$ yr$^{-1}$, with a median chirp mass of $15-25 M_{\odot}$.  Comparable rates of BHB mergers may be produced in globular clusters (e.g.,~\citealt{PortegiesZwart&McMillan02,O'Leary+06,Rodriguez+16,O'Leary+16}).  An alternative channel for producing tight BHBs via tidal locking of a massive main sequence binary and resulting homogeneous stellar evolution has recently been proposed \citep{Mandel&Demink16,Marchant+16}, which predicts a rate of $\sim 10$ Gpc$^{-3}$ yr$^{-1}$; a similar rate has been predicted at $z\approx 0$ from PopIII stellar remnant BH binaries \citep{Kinugawa+2014}.  The globular cluster and homogeneous evolution models, as well as the PopIII remnant binaries, exhibit a preference for more massive and equal mass binaries, similar to  GW150914.

\begin{figure} 
\begin{center}
\includegraphics[width=0.45\textwidth]{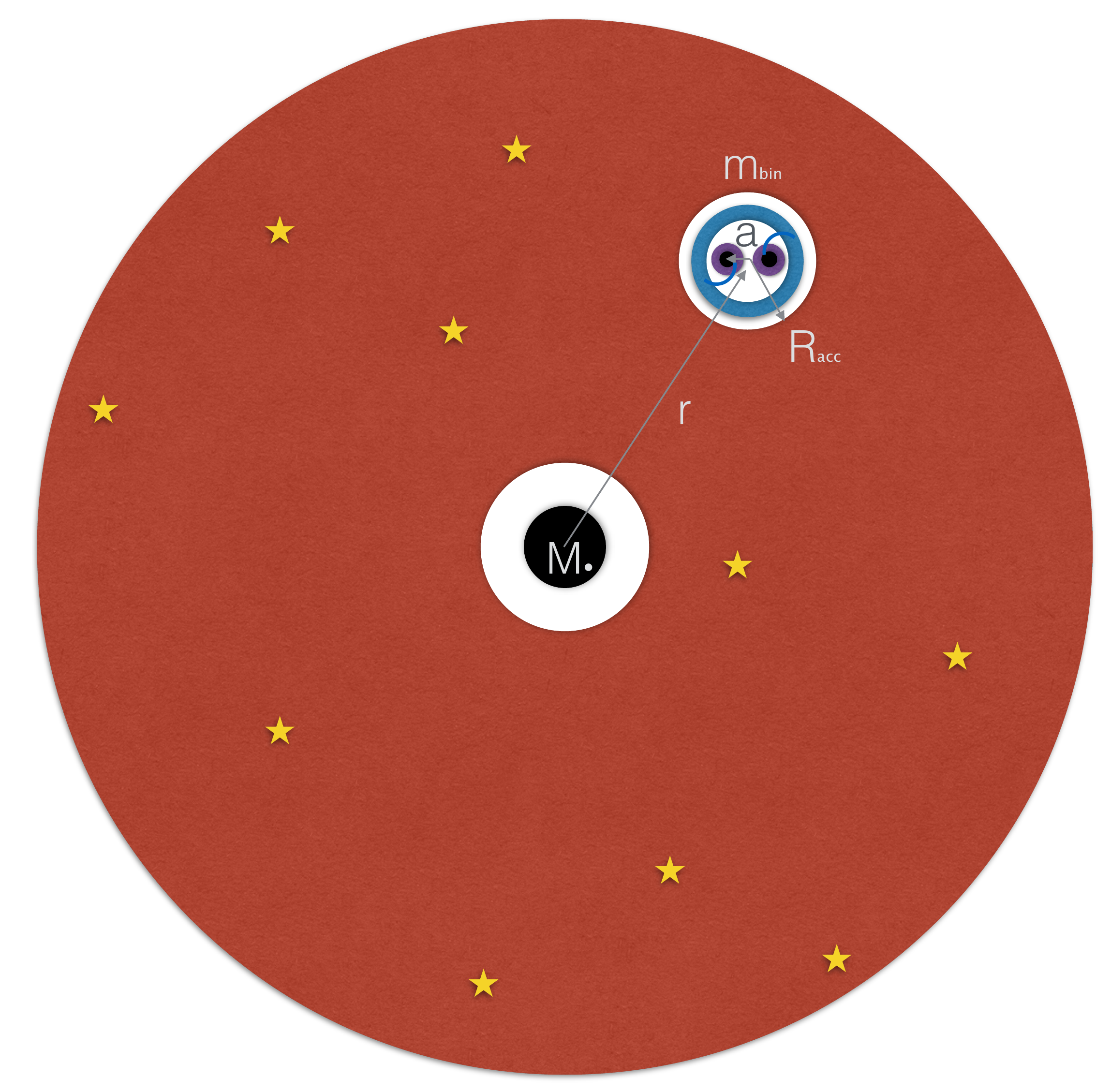}
\end{center}
\caption{A cartoon picture of our scenario.  A stellar mass black hole binary of mass $m_{\rm bin}$ orbits an SMBH of mass $M_\bullet$ at a distance $r$, embedded within a Toomre-unstable AGN accretion disk (red).  The AGN disk forms a disk of stars, some of which are binaries that evolve off the main sequence into BHBs.  A massive BHB will generally fail to clear a gap in the disk, but gas that flows into its accretion radius $R_{\rm acc}$ will form a circumbinary disk (blue) that hardens the binary's internal semimajor axis $a$.  Three-body interactions with background stars in the stellar disk (yellow) will also harden the BHB.  Gas streams peeling off the inner edge of the circumbinary disk can form minidisks around each BHB component (purple).}
\label{fig:cartoon}
\end{figure}

Here we propose a new channel for producing mergers of BHBs and other compact objects: the ``assisted inspiral'' of a stellar mass binary that is embedded in an active galactic nucleus (AGN) accretion disk.  Such disks are vulnerable to the Toomre self-gravitational instability (e.g.~\citealt{Goodman03}), and can form large quantities of stars during their lifetimes over a radial scale of parsecs.  Binary stars which form {\it in situ} will interact with the two--fluid disk (stars and gas) and in many cases can be helped along into a GW-driven inspiral, via interactions with one or both of these fluids.  A cartoon picture of our scenario is shown in Fig. \ref{fig:cartoon}.

Electromagnetic (EM) counterparts have been proposed for compact object mergers involving neutron stars (e.g., \citealt{Metzger&Berger12}).  However, no clearly promising counterparts have yet been identified for stellar mass BHB mergers because most emission scenarios require large quantities of surrounding baryonic matter, which is not expected for BHB mergers in the field or globular clusters.  The occurrence of a BHB merger in a gas-rich AGN disk makes conceivable a transient EM counterpart, produced by super-Eddington accretion onto the black hole merger product.  Less speculatively, the frequent association of this source class with strong AGN, a relatively rare galaxy type, will greatly reduce the number of possible host galaxies in the LIGO error volume.

In \S \ref{sec:diskModel} we describe the model we adopted for a self-gravitating and star-forming disk.  In \S \ref{sec:hardening}, we describe the various physical processes by which the internal orbit of a disk-embedded binary can evolve.  In \S \ref{sec:rates}, we provide preliminary estimates for the volumetric rates of BHB mergers produced in AGN disks, as well as the rate of such events detectable by Advanced LIGO.  In \S \ref{sec:observables}, we discuss potentially unique observational signatures of BHB inspirals produced through this novel channel, and we offer concluding remarks in \S \ref{sec:conclusions}.

%\newpage
\section{Disk Models}
\label{sec:diskModel}

\subsection{Gaseous Disk Model}

We employ the AGN accretion disk models of \citet{Thompson+05} (hereafter T05) for a supermassive BH (SMBH) mass $M_{\bullet} = 3\times 10^{6}M_{\odot}$, similar to those that contribute a large fraction of black hole growth in the local Universe (e.g.~\citealt{Hopkins+2007,Gallo+10}).  These models connect an  inner $\alpha$-disk, which is stable against gravitational fragmentation (Toomre parameter $Q > 1$), to an outer disk where star formation feedback by radiation pressure and supernovae is assumed to regulate the disk to a state of marginal gravitational stability $Q \approx 1$.  The amount of feedback required to support the disk self-consistently determines the star formation rate at each radius.  

The outer boundary condition is the gas feeding rate at $R_{\rm out} = $ 10 pc, which we take to be $\dot{M}(R_{\rm out}) = 15 \dot{M}_{\rm Edd}$, where $\dot{M}_{\rm Edd} = L_{\rm Edd}/0.1c^{2}$ is the Eddington accretion rate.  Additional parameters of the model assume fiducial values, including the viscosity parameter $\alpha = 0.1$; fraction $\epsilon = 10^{-3}$ of stellar rest mass placed into radiation; supernova feedback parameter $\xi = 1$; radial Mach number $\mathcal{M} = 0.1$ of the disk at radii where gravitational instabilities operate; and stellar velocity dispersion $\sigma = 180(M_{\bullet}/2\times 10^{8}M_{\odot})^{0.23}$ km s$^{-1}$ (\citealt{Kormendy&Ho13}).  

The T05 model for AGN disks involves a number of free parameters and assumptions.  The most important for our purposes are the location of $R_{\rm out}$ and the conditions (specifically, $\dot{M}$) there.  Observations of nearby AGN suggest that realistic disks extend to distances $0.1~{\rm pc} \lesssim R_{\rm out} \lesssim 10~{\rm pc}$ \citep{Burtscher+13}, motivating our fiducial choice of $R_{\rm out}=10~{\rm pc}$: if the reader is curious about disks truncated at smaller radii, our results can be simply cut off beyond the desired radius (disk conditions in the T05 model do not depend on conditions exterior to the radius of interest).  We note that our energy feedback parameter $\epsilon = 10^{-3}$ is a reasonable choice for nuclear energy release, but may underestimate the true feedback rate if accretion onto stellar mass BHs becomes the dominant source of heating.

\begin{figure} 
\begin{center}
\includegraphics[width=0.45\textwidth] {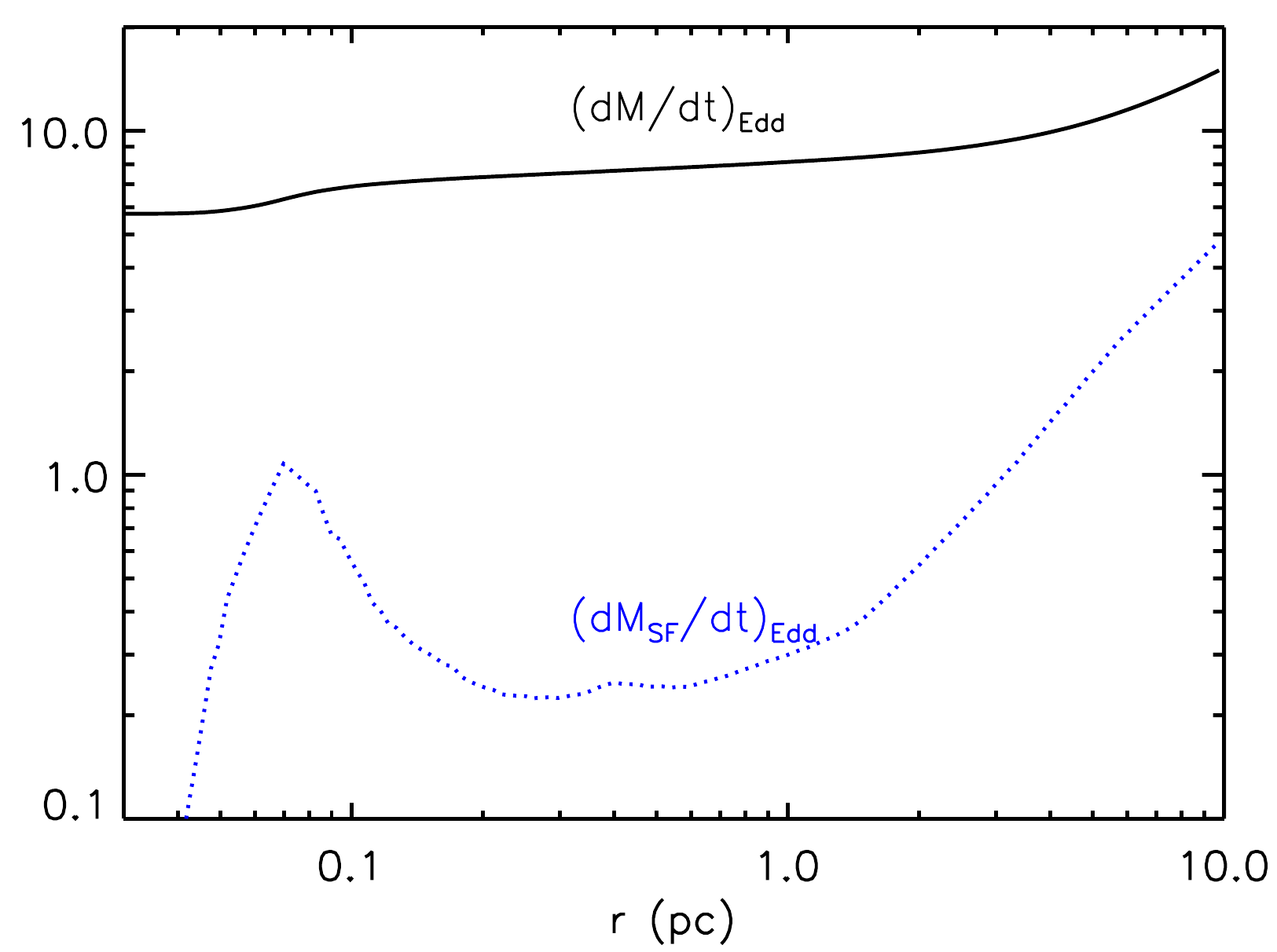}
\includegraphics[width=0.45\textwidth] {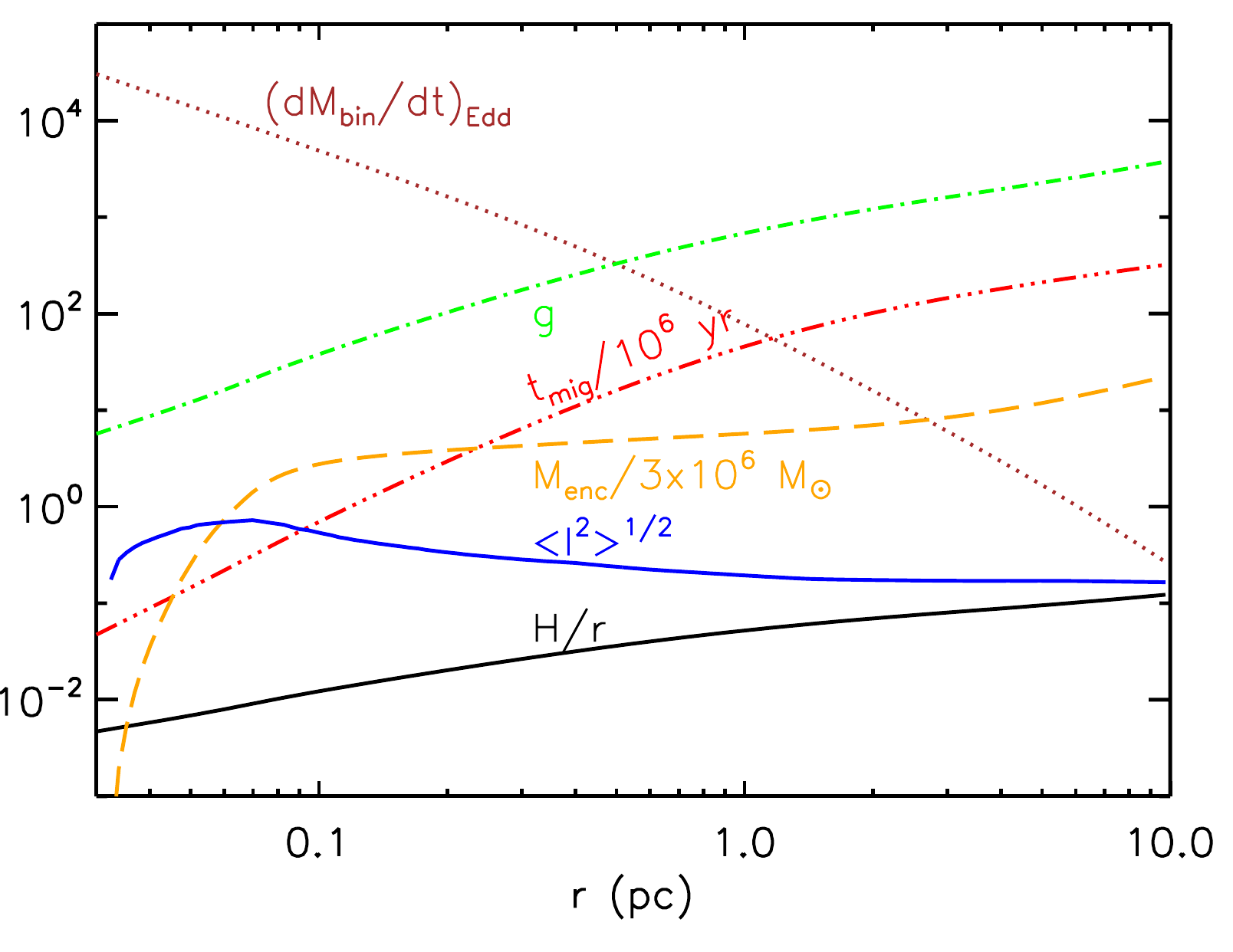}
\end{center}
\caption{Key quantities as a function of radius in the AGN disk around a $3\times 10^6M_\odot$ SMBH.  {\it Top Panel:} Gas accretion rate (black solid line) and star formation rate (dashed blue line), both in units of $\dot{M}_{\rm Edd}$.  Star formation is peaked near the T05 ``opacity gap'' at $\sim 0.1~{\rm pc}$ and also near the outer edge of the disk.  {\it Bottom Panel:} Disk aspect ratio $H/r$ (solid black line); binary gap opening parameter $g$ (dot-dashed green; a value of $g > 1$ indicates that the binary will not open a gap).  Accretion rate onto binary in Eddington units for the binary mass (dotted brown line).  Enclosed stellar mass formed over $T_{\rm AGN} = 10^{8}$ yr in units of the SMBH mass (orange dashed line).  Radial migration time of binary in Myr (triple dot dashed line).  Inclination dispersion of stellar disk due to two body relaxation after $T_{\rm AGN} = 10^{8}$ yr of evolution (solid blue line).}
\label{fig:disk}
\end{figure}

The top panel of Figure \ref{fig:disk} shows the mass accretion rate $\dot{M}$ and the star formation rate $\dot{M}_{\rm SF} = \pi r^{2}\dot{\Sigma}_{\star}$, where $\dot{\Sigma}_{\star}$ is the star formation rate per unit surface area.  About half of the stars form in a ring around 0.1 pc, with a comparable fraction forming at larger radii $\gtrsim 1-10$ pc.  Both regions correspond to locations where the opacity is low, requiring a higher rate of star formation to support the disk against self-gravity.  

The bottom panel of Figure \ref{fig:disk} shows other properties of the disk, such as the aspect ratio $H/r$, where $H$ is the vertical scale height of the gaseous disk, and the enclosed stellar mass $M_{\rm enc}(r)$.  The latter is calculated assuming that the steady-state star formation profile shown in the top panel is maintained for the typical AGN lifetime of $T_{\rm AGN} = 10^{8}$ years.  

Stars or BHBs of mass $m_{\rm bin}$ can migrate radially in the
gaseous disk.  The migration regime depends on whether the binary is
sufficiently massive to open a gap.  The criterion for gap-opening is
given by $g \lesssim 1$, where (\citealt{Lin&Papaloizou93};
\citealt{Baruteau+11})
\be
g \equiv \frac{3}{4}\frac{H}{r}\left(\frac{q}{3}\right)^{-1/3} + \frac{50\alpha }{q}\left(\frac{H}{r}\right)^{2}.
\ee 
Here $q = m_{\rm bin}/M_{\bullet}$, and $\Omega$ is the angular
velocity of the disk.  Figure \ref{fig:disk} shows that for a binary
of mass $m_{\rm bin} = 60M_{\odot}$, $g \gtrsim 1$ across all radii of
interest, so the Type I migration regime applies\footnote{We note that gaps can open if $\alpha \lesssim 10^{-2}$, as in \citet{Baruteau+11}.}.  The Type I
migration timescale $t_{\rm mig} \sim 1-100$ Myr across radii $r \sim
0.1-10$ pc, suggests that binaries might migrate towards the central
SMBH during the AGN lifetime.  However, these rates are very sensitive
to the thermodynamics of the gas near the binary, and are sufficiently
uncertain (e.g.,~\citealt{Paardekooper&Mellema06}) that it is
difficult to come to a definitive conclusion.

We estimate the gaseous accretion rate onto the BHB as
\be
\dot{M}_{\rm bin} = \pi \rho \sigma_{\rm gas} R_{\rm acc}\text{min}[R_{\rm acc},H],
\label{eq:Mdotbin}
\ee
where $c_{s}$ is the midplane sound speed, $R_{\rm H} \equiv a(m_{\rm bin}/M_\bullet)^{1/3}$ is the Hill radius for a binary orbiting the SMBH with semimajor axis $a$, and $\rho$ is the midplane gas density of the AGN disk.  The accretion radius and gas relative velocity are given by
\be
R_{\rm acc} = \frac{Gm_{\rm bin}}{\sigma_{\rm gas}^{2}},\,\,\, 
\sigma_{\rm gas} = [c_{s}^{2} + R_{\rm H}^{2}\Omega^{2} + (\chi r \Omega)^{2}]^{1/2},
\ee
respectively.  The three terms in $\sigma_{\rm gas}$ account for random hydrodynamic motions within the disk, the velocity shear across the finite extent of the Hill radius, and the relative velocity between gas and the BHB produced by noncircularity of the BHB orbit around the SMBH (we define $\chi$ in \S \ref{sec:hardening}).  At large distances ($\gtrsim 1~{\rm pc}$) from the SMBH, gas thermal velocities $c_{\rm s}$ dominate the denominator of $R_{\rm acc}$, but at smaller distances ($\lesssim 1~{\rm pc}$), it is the relative velocity between gas and orbit ($\chi r \Omega$) that controls the accretion rate onto stellar-mass objects.

Figure \ref{fig:disk} shows that $\dot{M}_{\rm bin}$ decreases from $\sim 10^{5}\dot{M}_{\rm edd}$ at 0.1 pc to $\sim \dot{M}_{\rm edd}$ at 10 pc.   At $\lsim 0.01$pc, the accretion rate onto the binary approaches the rate in the background AGN disk.  
These high accretion rates can drive the binary together through gaseous torques, and may provide a potentially luminous EM counterpart to the GW event.

\subsection{Stellar Disk Model}
We assume that stars forming in a Toomre-unstable disk will inherit the properties of the disk, meaning that their initial orbital eccentricities and inclinations will be $E_0 \sim c_{\rm s}^2/(2r^{2}\Omega^2)$ and $I_0 \sim H/r$, respectively.  For simplicity, we assume a Keplerian potential in this section, which is roughly correct for the radial scales of interest, and consider a stellar disk comprised of single-mass ($m$) stars unless otherwise noted.

As a dense collisional system, the evolving stellar disk will undergo internal two-body relaxation and eventually arrive at a more isotropic state \citep{Alexander+07}.  Collisional disks exist in two relaxational regimes distinguished by the origin of relative velocity between stars \citep{Rafikov&Slepian10}; they are shear-dominated if $\langle E^2 \rangle + \langle I^2 \rangle \lesssim (2m/M_\bullet)^{2/3}$ and dispersion-dominated if $\langle E^2 \rangle + \langle I^2 \rangle \gtrsim (2m/M_\bullet)^{2/3}$ (here $\langle E^2 \rangle^{1/2}$ and $\langle I^2 \rangle^{1/2}$ are the root mean squared eccentricity and inclination in the disk).  Because the model of T05 predicts relatively large $H/r$ values (Fig.~\ref{fig:disk}), we assume dispersion-dominated disks for the remainder of this paper.  In this regime, orbital eccentricities $E$ and inclinations $I$ follow a Rayleigh distribution \citep{Ida&Makino92}:
\begin{equation}
F(E^2, I^2)=\frac{1}{\langle E^2 \rangle \langle I^2 \rangle} \exp\left(-\frac{E^2}{\langle E^2 \rangle }-\frac{I^2}{\langle I^2 \rangle}  \right).
\end{equation}
The exponential suppression of highly inclined and eccentric orbits means that we can approximate all stellar orbits around the SMBH as almost circular unless $\langle E^2 \rangle \sim 1$ or $\langle I^2 \rangle \sim 1$.

\citet{Stewart&Ida93} find that a collisional disk with stars of a single mass will relax in a dynamical equilibrium where $\beta^2 \equiv \langle I^2 \rangle / \langle E^2 \rangle \approx 0.2$, and the time required to inflate a disk with local surface density $\Sigma_\star (r)$ is roughly
\begin{align}
T_{\rm I} &= \frac{2\pi^{1/2}M_\bullet^2\langle E^2 \rangle^{1/2} \langle I^2 \rangle^{3/2}}{\Sigma_\star \Omega r^2m BJ_{\rm z}}.
\end{align}
Here $B$ and $J_{\rm z}$ are dimensionless functions ($J_{\rm z}=2.7$ in $\beta$ equilibrium, and $20 \lesssim B \lesssim 30$).  We crudely approximate a typical (time-averaged) inclination spread for the stellar disk by equating $T_{\rm I} = T_{\rm AGN}$, resulting in
\begin{align}
\langle \bar{I}^2 \rangle^{1/2} \approx &0.23 \left(\frac{T_{\rm AGN}}{10^8~{\rm yr}} \right)^{1/4} \left( \frac{M_\bullet}{10^6 M_\odot} \right)^{-3/8} \\
&\times \left( \frac{\Sigma_\star}{10^5M_\odot~{\rm pc}^{-2}} \right)^{1/4} \left( \frac{m}{M_\odot} \right)^{1/4} \left( \frac{r}{\rm pc} \right)^{1/8}. \notag
\end{align}
The stellar disk thus puffs up to a vertical scale height $H_{\star} \sim r \langle \bar{I}^2 \rangle^{1/2}$, which is typically an order of magnitude larger than that of the gaseous disk (see Fig.~\ref{fig:disk}).  While most low-mass stars will quickly move into a thicker stellar disk, we do not expect BHBs or their progenitors to do so because of the evolution towards equipartition of random kinetic energy in multicomponent collisional systems \citep{Stewart&Ida93}.  If we consider a two-component stellar disk made primarily of low-mass stars with mass $m$ but with a small number of high-mass objects (e.g. BHBs) with mass $m_{\rm bin}$, equipartition would give the heavier objects a mean eccentricity and inclination of $\langle E_{\rm bin}^2 \rangle^{1/2} \sim \langle E^2 \rangle^{1/2}(m/m_{\rm bin})^{1/2}$, and $\langle I_{\rm bin}^2 \rangle^{1/2} \sim \langle I^2 \rangle^{1/2}(m/m_{\rm bin})^{1/2}$, respectively.  The relative velocity between these heavy objects and the background gas disk would then be $\chi r \Omega$, where $\chi = (\langle E^2 \rangle^{1/2}/2 + \langle I^2 \rangle^{1/2})(m/m_{\rm bin})^{1/2}$. 

In the above we have ignored interactions with pre-existing bulge stars.  Although these can be more numerous, their lower densities and higher relative velocities cause them to be less important for the relaxational evolution of the stellar disk.

\section{Hardening of Embedded Binaries}
\label{sec:hardening}

We now consider a binary of stellar mass objects with masses $m_1$ and $m_2$, and total mass $m_{\rm bin}=m_1 + m_2$, that are bound to an SMBH of mass $M_\bullet$ on an orbit embedded within the gaseous disk.  The internal semimajor axis of the binary is $a < R_{\rm H}$ to avoid tidal separation.  The primary focus of this paper is on binaries that form {\it in situ} in this disk, but we also note that pre-existing bulge stars can be captured into embedded orbits through hydrodynamical interactions (e.g. \citealt{Syer&Ulmer91}).  This latter process is inefficient and takes longer than a Hubble time in the disk regions considered in this paper, but can provide an additional source of disk-embedded binaries in denser, smaller scale AGN disks \citep{Bartos+16}.  In this section, we describe three physical processes that govern the orbital evolution of these binaries: gravitational radiation, interactions with single stars in the stellar disk, and interactions with the disk gas itself.

The GW merger timescale for a binary of point particles is, in the quadrupole approximation \citep{Peters64},
\begin{equation}
T_{\rm GW} = \frac{5a^4c^5}{256G^3 m_1 m_2 (m_1 + m_2)},
\end{equation}
where $a$ is the internal binary semimajor axis and we have assumed a circular orbit.  Because this timescale only becomes less than a Hubble time $T_{\rm H}$ for very tight orbits, we must search for additional mechanisms that can assist with the initial stage of merging BHBs.

In dense stellar systems, hard binaries will statistically lose energy to encounters with field stars, while soft binaries will gain energy and eventually dissociate \citep{Heggie75}.  The critical semimajor axis dividing these two regimes is
\begin{equation}
a_{\rm dis} = \frac{\lambda^2Gm_1 m_2 (m_1 + m_2 + m_3)}{m_3(m_1 + m_2)v_{\rm \infty}^2}.
\end{equation}
Here $m_3$ is the mass of the perturber star, $\lambda\approx 0.6$ is a dimensionless number that varies weakly with binary eccentricity, and $v_{\infty}$ is the relative velocity at infinity between the binary center of mass and the perturber.  For stars forming {\it in situ} in the AGN disk, we can approximate $v_{\infty} \sim \langle I^2 \rangle^{1/2}r\Omega$.

If we assume for simplicity that $m_1 = m_2$, then the evolution time $T_{\rm hard,\star} \equiv a({\rm d}a/{\rm d}t)^{-1}$ for a binary to significantly harden or soften is \citep{Spitzer87} 
\begin{equation}
T_{\rm hard, \star} = \frac{m_1 v_\infty}{2\pi K G n_\star m_3 (2m_1 + m_3)a},
\end{equation}
where $n_\star \equiv \Sigma_\star/ (r \langle \bar{I}^2 \rangle^{1/2})$ is the background number density of perturber stars and $K \approx 0.4$ is a dimensionless factor calibrated from numerical scattering experiments.  The value of $\Sigma_{\star} = \dot{\Sigma}_{\star}T_{\rm AGN}$ is determined by the star formation rate from our AGN disk solutions acting over the AGN lifetime $T_{\rm AGN}$ (Fig.~\ref{fig:disk}).
A soft binary will quickly run away to dissociation, but a hard binary will only harden more gradually, as its $T_{\rm hard,\star}$ grows with shrinking $a$.

The internal orbit of the binary may also evolve due to its interactions with ambient gas in the AGN disk.  Gas passing within the Hill sphere will be captured by the binary at the rate given by equation (\ref{eq:Mdotbin}; see also Fig.~\ref{fig:disk}).  Initially, the gas-driven hardening of the binary will be dominated by a complicated variant of gas dynamical friction (GDF).  The gravitational influence of the binary will create an overdense wake of disk gas behind each binary component that acts to decelerate it.  This process differs from the standard picture of GDF \citep{Ostriker99} in that (i) the wakes are necessarily nonlinear overdensities, and (ii) the binary is on a circular orbit, so the wakes wrap around the binary in a tadpole-like way.  

The nonlinear hydrodynamics of this hardening process have been simulated by \citep{Baruteau+11}, for a 15-15 ${\rm M_\odot}$ stellar binary orbiting in a thin disk around a $3\times10^6{\rm M_\odot}$ SMBH, similar to the systems envisioned here.  They find a circumbinary minidisk system, which is fed along narrow streams from gas crossing the gap created by the binary in the AGN disk.  The binary and its own minidisk is essentially a self-contained, compact binary+disk system, smaller in size than the Hill radius, and decoupled from the SMBH and the AGN disk.

The nonlinear hydrodynamics of a near-equal-mass binary with a
circumbinary disk has been addressed by many hydrodynamical
simulations, beginning with \cite{Artymowicz&Lubow94} (see, e.g. 
 \citealt{Hayasaki+07,Cuadra&Armitage09,Roedig+11,Shi+12,Farris+14} for more recent examples).
These simulations show that a central cavity, rather than an annular
gap, is cleared in the disc, and outside this cavity, the disk is
strongly distorted (note that Baruteau et al. 2011 perform a
simulation of the global AGN disk, and do not resolve this central
cavity in the minidisk).  Snapshots of the disk torques have been
measured in a few of the above simulations, and suggest that the
orbital decay is efficient
\citep{MacFadyen&Milosavljevic08,Cuadra&Armitage09,Roedig+11,Dorazio+13}. Unfortunately,
these results remain unreliable, because they are based on
oversimplified thermodynamics, and do not follow the system for
sufficiently long timescales (over which a ``live'' binary would
significantly change its orbit, and its disk geometry would likewise
evolve).

Here we resort instead to a toy model to estimate the secondary-dominated Type II migration rate $T_{\rm hard, gas}$, which is roughly consistent with the above simulations.  We estimate this Type II migration time using the formulae in \citet{Haiman+09} for the secondary-dominated regime of gas-driven torques (their Eq.~26).  Formally, the expressions from \citet{Haiman+09} break down at large semi-major axes of the binary gaseous disk near the Hill radius because the gas temperature drops below the value for which Kramers opacity is appropriate, $T \lesssim 10^{4}$ K.  However, the binary disk will be irradiated by the surrounding AGN disk, raising its temperature and decreasing the viscous time (and hence the migration time).  The estimates of \citet{Haiman+09} thus provide a conservative upper limit on the gaseous migration time.  We cap the total gas flow through the disk at the local Eddington limit.  Although significant uncertainty in the gas hardening rate remains, our subsequent results are not overly sensitive to this assumption because the gas-driven hardening time is typically very short compared to the AGN lifetime for binaries with semi-major axes $a \lesssim R_{\rm H}$ of interest.

The interplay between GWs, three-body scatterings, and circumbinary disks is complex.  At very small scales, gravitational radiation always dominates the rate of orbital decay.  At the largest separations, three-body scatterings are generally the most efficient source of orbital energy loss, but these become less effective as the binary hardens.  Torques from the circumbinary gas disk can bridge the gap, as the bottleneck for gas-driven inspirals is at large rather than small radii (the opposite of three-body scattering).  

We expect a newborn binary of massive stars to quickly harden due to three-body scatterings, but the binary is unlikely to be driven to merger by scatterings alone.  In contrast, our circumbinary disk model predicts that these stars can easily be forced into merger on timescales shorter than a stellar evolution time $T_\star \approx 3\times 10^{6}$ yr.

\begin{figure*} 
\begin{center}
\includegraphics[width=0.95\textwidth] {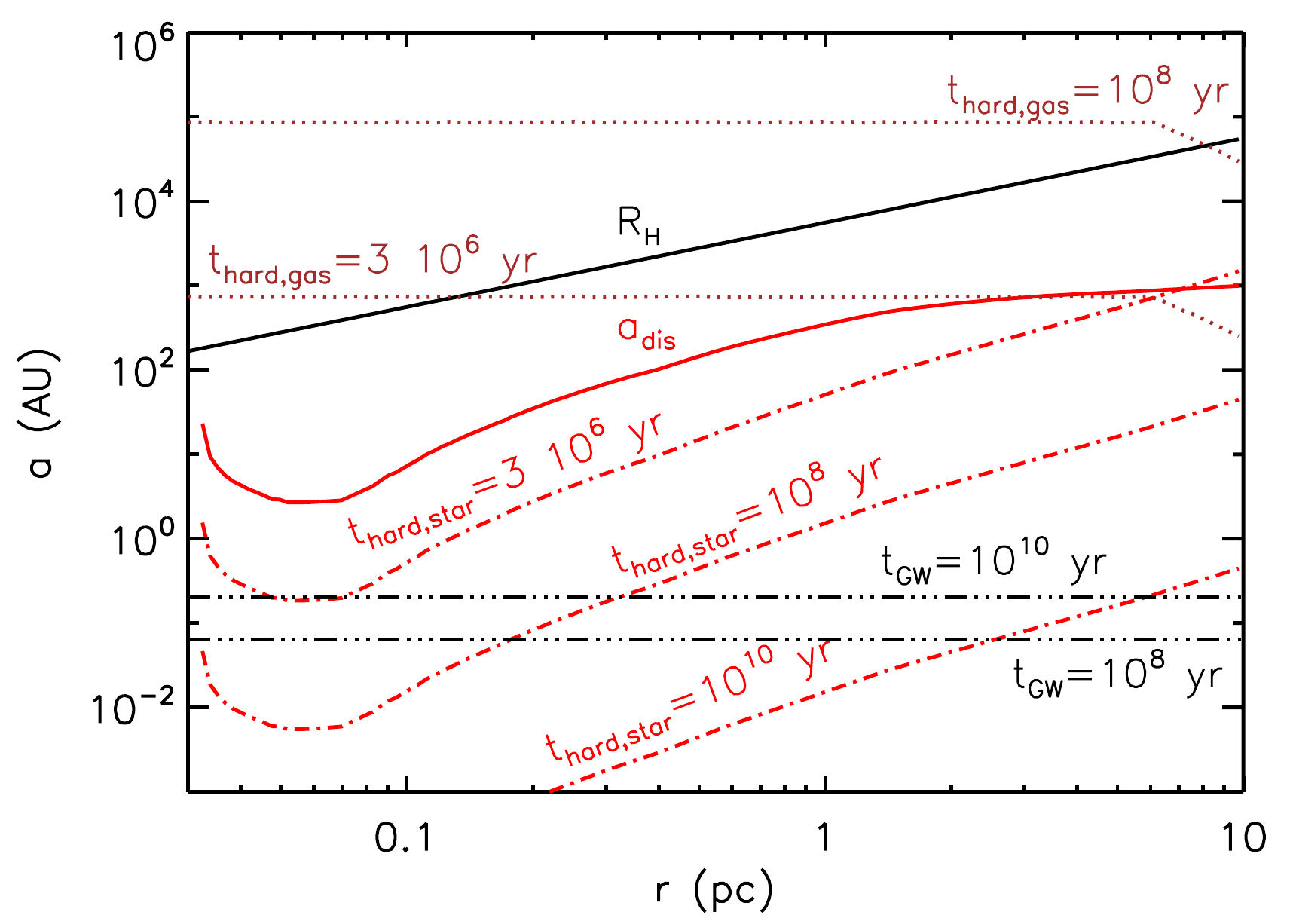}
\end{center}
\caption{Key values of the binary semi-major axis $a$ as a function of AGN disk radius $r$, including the Hill radius $R_{\rm H}$ (solid black line); the separation $a_{\rm dis}$ exterior to which binaries are dissociated through stellar collision (solid red line); the separation at which the GW inspiral time $t_{\rm GW} = 10^{8}, 10^{10}$ yr (horizontal triple-dot dashed lines).  The dotted brown lines show the semi-major axis distances interior to which the gas drag time is shorter than $3\times 10^{6}$ yr and $1\times 10^8$ yr.  The dot-dashed red lines show the separation exterior to which the stellar hardening time is shorter than $3\times 10^{6}$ yr, $10^{8}$ yr, and $10^{10}$ yr.  For the fiducial equal-mass binary with $m_{\rm bin}=60M_\odot$ shown here, both ``wet'' and ``dry'' mechanisms are often capable of hardening binaries of initially wide separation to semimajor axes $a<0.2~{\rm AU}$, where GW-driven coalescence occurs in a Hubble time and the ``final AU problem'' is solved.}
\label{fig:a}
\end{figure*}

Figure \ref{fig:a} shows various critical values of the binary semi-major axis as a function of radius in the AGN, calculated for our fiducial disk model (Fig.~\ref{fig:disk}).  The gas hardening time is found to be short ($\ll 3\times 10^{6}$ yr) at all semi-major axes $a<R_{\rm H}$ ($R_{\rm H}$ sets the outer edge of the binary gas disk) for stars forming near the opacity gap $r\sim 0.1~{\rm pc}$ .  However, the gas hardening time can be longer than $T_\star$ for the comparable number of stars forming at larger radii, near the disk outer edge.  

By contrast, the stellar hardening time is short compared to the stellar (AGN disk) lifetime outside of a binary separation marked as $T_{\rm hard,\star} = 3\times 10^{6}(10^{8})$ yr.  Notably, at radii $r \lesssim 1$ pc, these critical semi-major axes are interior to the separation $a \sim 200$ AU outside of which the stellar binary would be unbound if the BHs receive natal birth kicks of 50 km s$^{-1}$ (although birth kicks much smaller than this are possible for relatively large BHBs where most of the material in each supernova explosion eventually falls back, e.g. \citealt{Fryer+12}).  Despite considerably uncertainties in the gas hardening rate, Fig~\ref{fig:a} also shows that even once stellar hardening has ceased to be effective, gas hardening is more than sufficient to drive the binary separation to radii where GW emission can take over across a broad range of radii in the disk $\sim 0.1-10$ pc.            

One potential concern is that the massive stellar binary will be driven
to coalescence during the stellar lifetime itself, i.e. prior to BH
formation.  From Fig. \ref{fig:a}, it is clear that this only applies to stars formed near the opacity gap, not those formed in the outer regions of the AGN disk.  However, strong radiative and kinetic feedback from the O
stars could well stave off accretion in the binary by driving an
outflow, even up to infall rates of $\dot{M}_{\rm bin} \approx
10^{-3}M_{\odot}$ yr$^{-1} \approx 10^{3}\dot{M}_{\rm Edd}$
(\citealt{Wolfire87}).  While the energy and momentum of the O-star
winds are sufficient to halt accretion, whether this occurs in
practice, and the details of the process are not well understood.  In
particular, radiation pressure on dust could blow away the accreting
gas, especially given the high dust opacities expected in the AGN disk
on the radial scales of interest, although accretion may still occur
in the plane of a 2D disk and/or along unstable thick filaments
(e.g. \citealt{Krumholz+2009}).  However, a pressurized compact HII
region and photoevaporation in the innermost regions of such a disk
could result in starving the massive star entirely
\citep{Hosokawa+2011}.

To make this argument more concrete, consider the impact of main sequence O star winds on unbound AGN disk gas that is approaching the binary with a pericenter $R<R_{\rm acc}$.  Absent feedback, this material will join a circumbinary accretion disk following Bondi-Hoyle accretion.  However, as it approaches the binary, it will be blown out by ram pressure of the wind if $\dot{M}_{\rm w} v_{\rm w}^2 \gtrsim \dot{M}_{\rm bin} Gm_{\rm bin}/R$.  Modern theoretical calculations \citep{Muijres+12} predict that O stars on the main sequence experience mass loss rates $\dot{M}_{\rm w} \sim 10^{-6}M_\odot~{\rm yr}^{-1}$ with terminal wind speeds $v_{\rm w} \sim 3000~{\rm km~s}^{-1}$.  The enormous ram pressure delivered by these winds can overwhelm inflowing material during the Bondi-Hoyle stage of accretion, unbinding it before it has the chance to enter a circumbinary disk.  The ratio of wind ram pressure to the critical value needed for blowout is
\begin{equation}
\Upsilon \equiv \frac{\dot{M}_{\rm w}v_{\rm w}^2}{\dot{M}_{\rm bin}Gm_{\rm bin}/R_{\rm acc}}.
\end{equation}
For our fiducial binary and disk parameters, $\Upsilon \sim 4 $ ($\Upsilon \sim 4000$)  at $r=0.1~{\rm pc}$ ($r = 1~{\rm pc}$), indicating that O star winds will strongly suppress circumbinary accretion torques except at the smallest radii from the SMBH.

Because O star winds are roughly isotropic, and the winds and jets from accreting black holes are focused along their polar axes, we expect ram pressure from BHB outflows to only unbind a fraction of the incoming gas.  Even if BHB outflows unbind a large majority of AGN gas approaching them, the BHBs only generate feedback while actively accreting, implying a limit cycle behavior that may nonetheless permit BHBs to be hardened through circumbinary torques (though with less efficiency than Fig. \ref{fig:disk} predicts).

Finally, we note that black hole birth kicks have the potential to unbind some fraction of O star binaries formed in the outer regions of the star forming disk (but not those forming near the opacity gap, where stellar hardening brings the binary to a tight separation on a stellar evolution timescale).  The O star binaries in this scenario may not be favorable for forming double neutron star systems, however, as the usual problem of unbinding due to mass loss still occurs \citep{Brandt+95}.

\section{Rates}
\label{sec:rates}

The mass density of SMBHs in the local universe ($z\approx 0$) is estimated to be $\rho_{\bullet} \approx 4\times 10^{5}M_{\odot}$ Mpc$^{-3}$ \citep{Yu&Tremaine02, Shankar+04}.  Because most BHB mergers catalyzed by AGN disks occur on timescales $\ll$ Gyr, we need only focus on AGN disks out to the LIGO detection horizon, which we approximate as $z \approx 0$ for the purposes of estimating SMBH growth histories.  Specifically, $\dot{\rho}_{\bullet}(z\approx0) \approx  3\times 10^{-6}M_{\odot}$ Mpc$^{-3}\,$yr$^{-1}$ \citep{Marconi+04}.

Nonrotating stars of zero-age main sequence mass $M_{\star} \gtrsim 20M_{\odot}$ are expected to form black holes following the end of their nuclear-burning lives (e.g., \citealt{Vink+01, Heger+03}).  For a standard Kroupa IMF \citep{Kroupa01}, only a fraction $f_{\bullet,n} \approx 2\times 10^{-3}$ of the stars (by number) are born with $\gtrsim 20M_{\odot}$, but these account for a fraction $f_{\bullet} \approx 10^{-1}$ of the total star forming mass.  On the other hand, it is unclear whether star formation in self-gravitating disks obeys the same IMF seen in field stars.  The stellar disk in the central parsec of the Milky Way is one of our only probes of disk-mode star formation (\citealt{Levin&Beloborodov03}), and there a top heavy IMF with ${\rm d}N/{\rm d}M_\star \propto (1/M_{\star})^{1.35-1.7}$ \citep{Paumard+06, Alexander+07, Lu+13} - or an even more radical IMF with ${\rm d}N/{\rm d}M_\star \propto M_\star^{-0.45}$ \citep{Bartko+10} - may be consistent with the observed population.  Such top-heavy IMFs are further motivated by numerical simulations of star formation in marginally stable AGN disks \citep{Nayakshin+07}.  In this case $>20M_{\odot}$ stars account for a fraction $f_{\bullet} \gtrsim 0.5$ by mass.  Note that the lifetime of $> 20M_{\odot}$ stars prior to core collapse is also $\lesssim 10^{7}$ years, i.e. a typical AGN duty cycle.  

For every gram of mass accreted by the SMBH, we assume that a fraction $f_{\star}$ is processed through a self-gravitating disk into stars, i.e. the total star formation rate in self-gravitating AGN disks is $\dot{\rho}_{\star} = f_{\star} \dot{\rho}_{\bullet}$.  Based on the disk models of T05, a value of $f_{\star} \gtrsim 1$ is likely, due to the minimal amount of star formation is required to support the disk through the opacity gap.  Although in principle values of $f_{\star} \gg 1$ are achievable, the $\sim ~{\rm pc}$ scale size of the opacity gap (which is insensitive to most parameters of the model) means that $f_\star \gg 1$ would greatly overproduce the central starlight of typical galaxies.  This constraint is particularly severe for larger galaxies: star formation through the T05 model will produce a central stellar density $\sim f_\star M_\bullet~{\rm pc}^{-3}$, and an SMBH influence radius $r_{\rm inf} \sim f_\star^{-1/3}{\rm pc}$.  This is only compatible with the observed influence radii of smaller, $M_\bullet \sim 10^6 M_\odot$, SMBHs (for example, \citealt{Stone&Metzger16} fit $r_{\rm inf} = 16~{\rm pc}(M_\bullet/10^8M_\odot)^{0.69}$ to a large sample of nearby galaxies). 

However, the observed ``cosmic downsizing'' trend \citep{Gallo+10} means that most of the local universe's $\dot{\rho}_\bullet$ is concentrated in smaller SMBHs with $10^6 M_\odot \lesssim M_\bullet \lesssim 10^8 M_\odot$.  These SMBHs are indeed surrounded by dense nuclear star clusters \citep[NSCs;][]{Georgiev&Boker14} which can in many cases dominate the mass of the central parsecs \citep{Georgiev+16}.  The origin of these NSCs is debated, and some of their mass likely arrives in the form of inspiralling globular clusters \citep{Tremaine+75, Leigh+12, Gnedin+14}, but observations increasingly suggest that a majority of their mass has likely formed {\it in situ} \citep{Antonini+15, Leigh+15}.  This motivates our fiducial choice of $f_\star =1$, as standard star formation mechanisms should be suppressed by the SMBH tidal field at such small scales.  We also note that if disk-mode star formation involves a top-heavy IMF, observational constraints limiting the maximum value of $f_\star$ become much weaker, as short-lived high-mass stars would dominate the total mass in the stellar disk\footnote{This would leave behind a disk of compact remnants, the dynamics of which would certainly be interesting but which also are not easily constrained by observed surface brightness profiles.}.

The field binary mass fraction for O stars is $f_{\rm bin} \simeq 0.69\pm 0.09 $ (\citealt{Sana+12}).  Assuming that Opik's law applies to high-mass stars \citep{Kobulnicky&Fryer07}, the cumulative distribution of primordial binary separations $a$ is logarithmically flat between a minimum value $a_{\rm min} \sim 0.01$ AU and a maximum value $a_{\rm max} \sim 10^{5}$ AU.  Truncating this distribution at the hard-soft boundary of $ \sim 10^{2}$ AU, we obtain a massive stellar binary natal fraction of $f_{\rm bin} = 0.56$.  This calculation of $f_{\rm bin}$ is motivated by observations of massive field binaries because of the much greater uncertainties concerning binarity in disk-mode star formation.  Theoretically, a high, order unity binary fraction is generally expected to accompany disk-mode star formation in galactic nuclei because of the relatively fast cooling times characteristic of these systems \citep{Alexander+08}.  Binary formation is also seen in simulations of brown dwarf formation in Toomre-unstable protostellar disks \citep{Stamatellos+07, Thies+10}.  Moreover, observations of O stars in the Galactic Center's stellar disk find a binary fraction comparable to that in young massive clusters \citep{Pfuhl+14}.  

In conclusion, if a fraction $f_{\rm m}$ of BHBs produced in the disk merge, the local volumetric event rate is 
\be
\mathcal{R} \sim f_{\star}f_{\rm bin}f_{\bullet}f_{\rm m} \frac{\dot{\rho}_{\bullet}(z=0)}{\langle m_{\rm O}\rangle} \approx 60 f_{\star}f_{\rm bin}f_{\bullet}f_{\rm m}{\rm Gpc^{-3}}{\rm yr^{-1}}
\ee
where $\langle m_{\rm O} \rangle$ is the mean mass of the binary O star progenitor system, taken to be $\approx 50M_{\odot}$ in the final line.  The rate of BHB mergers through standard channels \citep[i.e. binary evolution of field stars:][]{Dominik+15} and dynamical formation scenarios \citep{O'Leary+06, O'Leary+09, Rodriguez+16} is highly uncertain, but the event rate of the formation channel proposed in this paper appears competitive with these scenarios and those consistent with the discovery of GW150914 \citep{LIGO+16b}, although more recent results from the first Advanced LIGO observing run have increased the estimated volumetric BHB merger rate to $9-240~{\rm Gpc}^{-3}~{\rm yr}^{-1}$ \citep{Abbott+16}.  In a fiducial model ($f_\star = 1$, $f_{\rm bin}=0.56$, $f_\bullet = 0.1$, $f_{\rm m}=1$), we expect a merger rate of $\mathcal{R} = 3~{\rm Gpc}^{-3}~{\rm yr}^{-1}$, but note that this could easily increase an order of magnitude with a top-heavy IMF, and emphasize that many other uncertainties have entered into this estimate.

\section{Observables}
\label{sec:observables}

\subsection{Distribution of Binary Parameters at Coalescence}
If there is a phase of orbital inspiral mediated by torques from a circumbinary disk, the minidisks forming around individual BHs (of mass $m_\bullet$, spin $a_\bullet$) will be able to reorient their spin vectors.  The Lense-Thirring effect couples a misaligned minidisk to the spin axis of the BH, torquing the BH and (given enough time and inflowing mass) eventually bringing spin and disk angular momentum into alignment.  There are two different physical limits of relevance.  In the first limit, the minidisk aspect ratio $h/r < \alpha$, the dimensionless Shakura-Sunyaev viscosity parameter of the minidisk.  In this regime, warps produced by differential nodal precession propagate in a diffusive manner and the disk aligns with the BH equatorial plane out to a Bardeen-Petterson radius \citep{Bardeen&Petterson75, Miller&Krolik13}
\begin{equation}
r_{\rm BP} \sim r_{\rm G} \left(\frac{2a_\bullet}{\alpha} \right)^{2/3} \left(\frac{h_{\rm BP}}{r_{\rm BP}} \right)^{-4/3},
\end{equation}
where $r_{\rm G}=Gm_\bullet/c^2$.  This regime is most relevant for radiatively efficient sub-Eddington accretion where the disk is well-described by the solution of \citet{Shakura&Sunyaev73}, i.e. BHBs formed in the outer regions of the AGN disk.  Generally, $r_{\rm BP}<R_{\rm acc}$.  Applying the Bardeen-Petterson alignment timescale \citep{Miller&Krolik13} to a \citet{Shakura&Sunyaev73} model for minidisks around individual components of the BHB yields spin alignment on timescales much shorter than the secondary-dominated Type II migration timescale of \citet{Haiman+09}, as expected by analogy to SMBH binaries \citep{Bogdanovic+07}.

In the opposite regime, $h/r > \alpha$ and warps propagate as bending waves in the minidisks.  Long-term alignment of BH spin is not well-studied in this regime, but the alignment times could be even shorter here because disk misalignment can persist down to the innermost stable circular orbit (though whether this occurs in practice depends on whether the minidisks are prograde or retrograde; see \citealt{Nealon+15}).  Overall, we expect rapid spin alignment unless BHB feedback completely chokes off accretion. 

Binaries created through disk-mode star formation are expected to form through dissipative capture rather than fission of collapsing clouds \citep{Alexander+08}, implying that orbits both prograde and retrograde with respect to the circumbinary disk should exist.  Retrograde binaries experience much faster initial rates of hardening (during the dynamical friction dominated early phases simulated by \citealt{Baruteau+11}).  Because both prograde and retrograde circumbinary disks with some initial tilt ultimately settle into an aligned or counteraligned configuration \citep{Nixon12}, we expect spin alignment to occur in both regimes.

For those binaries that do accrete significantly (increasing their
mass by a factor $\gtrsim 2$) before merger, we would expect a
convergence not just in spin orientation but also in spin magnitudes
and even mass ratio.  Circumbinary accretion into a cavity tends to
equalize the mass ratio in a binary \citep{Farris+14}, and order unity
mass growth of black holes will spin them up to theoretical maxima set
by details of the accretion flow.  For example, in a radiatively
efficient thin disk, \citet{Thorne74} found an asymptotic upper limit
on $a_{\rm BH}$ of $a_{\rm max}\approx 0.998$.  The exact value of
this upper limit is sensitive to accretion rate and effective
viscosity $\alpha$ but in most cases is $a_{\rm max} \gtrsim
0.95$ \citep[but see also references therein]{Sadowski+11}.  However,
our model for secondary-dominated Type II migration finds that BHBs
generally accrete only a small fraction of their mass: even though
they can accrete at super-Eddington rates, they also can harden much faster
than a Salpeter time. As a result, the masses, spin magnitudes, and
mass ratios of BHBs hardened in AGN disks will generally reflect the
birth distributions of these parameters for disk-mode star formation
(and stellar evolution).  For example, if GW151226 originated via the channel proposed here, the measured spin parameter $a_{\rm BH}>0.2$ (for one binary component) would likely be natal in origin.

\subsection{Transient Electromagnetic Counterparts}

The BHB formation channel identified in this paper is uniquely capable of producing a strong, transient electromagnetic (EM) counterpart; as far as we are aware it is the only known mechanism by which stellar mass black holes are driven to merger in the presence of substantial gas densities (the disk-embedded binaries envisioned here could alternatively arise from pre-existing binaries dragged into the disk; \citealt{Bartos+16}).  In this section, we offer preliminary estimates for the nature and detectability of these counterpart signals, but emphasize that they are highly uncertain and must be better quantified in future work.

Figure \ref{fig:disk} shows binary accretion rates of $\dot{M}_{\rm bin} \sim 1-10^{5}\dot{M}_{\rm Edd}$, depending on radius in the disk, the luminosity from which could provide a possible EM counterpart prior to and following the merger.  Highly super-Eddington accretion is potentially susceptible to powerful radiation-driven outflows, which reduce the fraction of the inflowing gas that ultimately reaches the central BH and its radiative efficiency $\eta$.  However, recent radiation magnetohydrodynamic simulations of accretion disk find that large efficiencies $\eta \gtrsim 0.04$ are achieved even for highly super-Eddington accretion (\citealt{Jiang+14,Sadowski&Narayan15}).  If a significant fraction of the bolometric luminosity $L_{\rm Bol} = \eta \dot{M}c^{2} \sim 10^{40}(\eta/0.1)(\dot{M}/\dot{M}_{\rm Edd})$ erg s$^{-1}$ emerges in the X-ray band (e.g., as in ultra-luminous X-ray sources), then the resulting X-ray luminosities up to $L_{\rm X} \sim 10^{44-45}$ erg s$^{-1}$ are comparable to those of the AGN, and readily dectable by {\it Chandra} or {\it Swift}.  

However, if the radiative efficiency is much lower $\eta \ll 1$ due,
e.g., to super-Eddington outflows, then $L_{\rm X}$ will be much lower
and the BHB will be challenging to detect relative to the more
luminous AGN.  Furthermore, the merged black hole will experience a birth kick $v_{\rm k}$ due to anisotropic emission of GWs \citep{Fitchett83}.  Only the portion of the circumbinary disk (or surviving minidisks) interior to $r_{\rm k} = Gm_{\rm bin}/v_{\rm k}^2$ will remain bound to the recoiling black hole \citep{Loeb07}, and any accretion luminosity will subside on the viscous time at this radius, $T_{\rm dim} = (Gm_{\rm bin}/v_{\rm k}^3)\alpha^{-1}(h_{\rm k}/r_{\rm k})^{-2}$, where $h_{\rm k}/r_{\rm k}$ is the surviving disk aspect ratio at the kick radius $r_{\rm k}$.  If we take fiducial values $h_{\rm k}/r_{\rm k}=0.5$, $\alpha=0.1$, $m_{\rm bin}=60 M_\odot$, and $v_{\rm k}=100~{\rm km~s}^{-1}$ (typical of comparable mass, spin aligned mergers as in \citealt{Lousto+10}), we get a dimming time $T_{\rm dim} \sim 10~{\rm yr}$, but this can be much shorter if spin alignment is only partial \citep{Lousto+12}.  Only for the largest kicks, $v_{\rm k} \gtrsim 10^3~{\rm km~s}^{-1}$, is the viscous time in the surviving disk short enough to seriously impede observability.

\subsection{Environmental Search Strategies}
For reasons discussed above, we expect the majority of these gas-induced mergers to occur during a phase of AGN activity, which will facilitate the search for host galaxies of GW signals.  Anticipated sky error regions for the LIGO-Virgo network are $\Delta \sim 10-100$ ${\rm deg}^2$, so if we assume an event is observed at luminosity distance $d_{\rm L}$ with distance uncertainty $\delta d_{\rm L}$ (note that typically, $\delta d_{\rm L} \sim d_{\rm L}$ because of a degeneracy between distance and inclination), then the total number of AGN in the error volume is
\begin{equation}
N_{\rm AGN} \approx 380 \frac{\phi_{\rm AGN}}{10^{-4}~{\rm Mpc}^{-3}} \frac{\Delta}{600~{\rm deg}^2}\left(\frac{d_{\rm L}}{400~{\rm Mpc}} \right)^3 ,
\end{equation}
where $\phi_{\rm AGN}$ is the approximate space density of AGN in the local universe.  For events similar to GW150914 with $d_{\rm L} = 400$ Mpc, $\delta d_{\rm L} = 350$ Mpc, and $\Delta = 600$ deg$^{-1}$, we expect $\sim 10^{2-3}$ AGN in the search volume.  $N_{\rm AGN}$ could be reduced much further for a well-localized event in the future with $\Delta = 10$ deg$^{2}$, as larger networks of detectors come online \citep{Nissanke+11}, or if measurement of BH spins breaks the degeneracy between distance and inclination.  
Also, above we have assumed for simplicity, assumed a constant comoving AGN density $\phi_{\rm AGN}$ and a Euclidean low-redshift universe.  Judicious cuts on galaxy luminosity can reduce the number of candidate hosts further \citep{Gehrels+15}. 

Reducing the candidate hosts to a modest number of galaxies will greatly increase the tractability of EM counterpart searches.  The notion of finding a unique quasar counterpart to a BH binary GW source was proposed in the context of SMBH mergers expected to be detected by LISA \citep{Kocsis+2006}; interestingly, a similar search for an EM counterpart among quasars appears feasible in our scenario for LIGO.

\section{Conclusions}
\label{sec:conclusions}
We have argued that Toomre-unstable AGN disks can form appreciable numbers of hard stellar mass BHBs, which will be driven to coalescence by a combination of (i) statistical hardening from three-body encounters with background disk stars, (ii) gaseous torques from a circumbinary minidisk, and (iii) GW emission once the BHBs have reached very small separations.  This novel BHB formation channel will (at low redshift) preferentially produce mergers in small galaxies with SMBH masses $10^6 M_\odot \lesssim M_\bullet \lesssim 10^8 M_\odot$.  As a proof of principle, we have used the star-forming AGN disk model of T05 to estimate rates of BHB production in a representative, $f_\star \approx 1$ disk around a $3\times 10^6 M_\odot$ SMBH.  In future work, we will conduct a wider parameter study across a range of SMBH masses and disk parameters.

There are many uncertainties involved in this BHB formation channel that must also be investigated further.  The most important assumptions we have made in producing our fiducial rate estimate of $\mathcal{R} \sim 3 ~{\rm Gpc}^{-3}~{\rm yr}^{-1}$ are the following:
\begin{itemize}
\item Feedback from O stars is capable of greatly limiting accretion onto an O star binary.  This is currently an unresolved question in the literature, but it is a crucial component of our model because of the ability of torques from a circumbinary minidisk to rapidly merge O stars on the main sequence, which would prevent the formation of a BHB.  
\item The question of feedback from a BHB onto a circumbinary minidisk is also an interesting one, but is of secondary importance because BHBs can generally be driven into the GW regime by three-body encounters with disk stars.  If gas torques drive BHBs into the GW regime, their spins will generally be aligned at merger, but this is not the case if they are brought together by stellar scatterings.
\item The fraction of stellar disk mass that goes into the progenitors of BHBs, $f_\bullet$, depends sensitively on the IMF for disk-mode star formation.  We have conservatively assumed a Kroupa IMF, but note that if the top-heavy disk-mode IMFs inferred by observations of the Galactic Center are more valid, then $\mathcal{R}$ could increase by an order of magnitude.  A top-heavy IMF would also give this channel the ability to further increase rates by loosening observational constraints on $f_\star$.
\end{itemize}

The T05 model itself is a steady-state simplification of the time-dependent physics of AGN activity.  In some hydrodynamical simulations of gaseous AGN disks, star formation occurs in rapid bursts, particularly when feedback is ineffective \citep{Nayakshin+07}.  In this limit the disk gas is quickly exhausted.  However, given the robustness of the stellar hardening mechanism (particularly in the inner disk), we would still expect a significant fraction of BHBs formed in this way to merge with external assistance.

To summarize, star formation in AGN disks appears to be a promising channel for producing BHB mergers in the local universe, but depends sensitively on the ability of feedback to limit circumbinary disk accretion, and on the IMF for star formation driven by the Toomre instability.  This BHB formation channel is particularly intriguing at the start of the era of GW astronomy, given indications that BHBs may dominate the rate of future LIGO detections.  While most BHB formation channels predict electromagnetically dark mergers that are difficult to localize on the sky, the mechanism we advance in this paper will generically merge BHBs in gas-rich environments and in a rare subset of galaxies, motivating EM followup searches.

\section*{Acknowledgments}
We thank Joshua Bloom, Saavik Ford, Kohei Inayoshi, Barry McKernan and Cole Miller for useful discussions.  BDM gratefully acknowledges support from NASA {\it Fermi} grant NNX14AQ68G,  NSF grant AST-1410950, and the Alfred P. Sloan Foundation.  Financial support was provided to NCS by NASA through Einstein Postdoctoral Fellowship Award Number PF5-160145 and to ZH by NASA ATP grants NNX11AE05G and NNX15AB19G.  ZH gratefully acknowledges support from a Simons Fellowship in Theoretical Physics.

\bibliographystyle{mn2e}
\bibliography{ms}

\end{document}